\documentclass[12pt,a4paper]{article}
\usepackage{graphics}
\usepackage{array}
\usepackage{hhline}
\usepackage{subfigure}
\usepackage{amssymb}
\usepackage{a4}
\usepackage{graphics}
\usepackage{array}
\usepackage{hhline}
\usepackage{subfigure}
\usepackage{euscript}
\newcommand{\be}{\begin{equation}}
\newcommand{\ee}{\end{equation}}
\newcommand{\bea}{\begin{eqnarray}}
\newcommand{\eea}{\end{eqnarray}}
\newcommand{\nn}{\nonumber\\}
\begin{document}
\title{Next-to-leading order vs. quark off-shellness
and intrinsic $k_T$ in the Drell-Yan process
\thanks{work supported by BMBF}}
\author{O.~Linnyk\thanks{olena.linnyk@theo.physik.uni-giessen.de}, S.~Leupold, U.~Mosel }
\date{\small Institut f\"ur Theoretische Physik, Universit\"at Giessen, Germany \\
 \today}
\maketitle
\begin{abstract}
We calculate the effects of next-to-leading order perturbative QCD
as well as of the quark transverse motion and off-shellness on the
Drell-Yan process cross section.  By studying the $s \to \infty$
behaviour of the cross section in these approaches, we find that
the effects of quark off-shellness and intrinsic-$k_T$ parametrize
those of higher twists. In particular, the off-shellness of
partons generates part of the $K$-factor type corrections to the
leading order cross section. Higher twist contributions to the
$p_T$-spectrum of the Drell-Yan pairs are found to be large for
presently accessible energies. The evolution of quark
off-shellness distribution with the hard scale is also studied.

PACS numbers:
13.85.Qk, 
12.38.Cy, 
12.38.Qk  
\end{abstract}


\section{Introduction} \label{intro}

At the future GSI-FAIR facility~\cite{FAIR}, several experiments
will perform precise measurements of high mass lepton pair
production in $\bar p p$ and $\bar p A$ collisions in order to
determine parton distributions in the proton and in nuclei.
Experiments GSI-PAX~\cite{PAX} and GSI-ASSIA~\cite{ASSIA} are
going to study the polarized process, while GSI-PANDA~\cite{PANDA}
will investigate unpolarized $\bar p p \to l^+ l^- X$ at $s$ as
low as 15~GeV$^2$.

A good theoretical understanding of the Drell-Yan process at the
level beyond the leading order (LO) of perturbative QCD (pQCD) is
necessary. Indeed, the effects beyond the LO pQCD are expected to
be high at this low energy~\cite{paper3,E866}. These effects on
the double differential cross section can be roughly parametrized
by an overall $K$-factor, giving the discrepancy between the LO
calculations and the data. However, in case of the triple
differential cross section ({\it i.e.} transverse momentum
distribution of dileptons), the corrections to LO pQCD cannot be
parametrized by a constant factor, as we discuss below.

Applied to the Drell-Yan process $\! \bar p p \to l^+ l^- X \!$,
the leading order approximation of collinear pQCD predicts the
correct dependence of the {\em double differential} Drell-Yan
cross section $d^2 \sigma/dM^2 d x_F$ on the hard scale $M$. Here
\be \label{xf} x_F\equiv p_z / |p_z|_{max} \ee
is the Feynman variable of the lepton pair and $M$ its invariant
mass. However, LO pQCD fails to reproduce
\begin{enumerate}
\item the magnitude of this cross section, the discrepancy being usually
parametrized by a $K$-factor;
\item the average transverse momentum $p_T$ of the dileptons;
\item the $p_T$-spectrum of Drell-Yan pairs, which is given by the {\em triple
differential} cross section $d^3 \sigma/dM^2 d x_F dp_T$.
\end{enumerate}
Experimentally observed Drell-Yan lepton pairs have non-vanishing
transverse momentum $p_T$, which can be as large as several GeV.
However, in leading order of pQCD, the cross section is
proportional to $\delta (p_T)$. Indeed, in collinear QCD, both the
transverse momentum and the light cone energy of quarks inside
hadrons are neglected compared to the large component of the quark
momentum, parallel to the hadron momentum. Thus, the initial
state, {\it i.e.} the colliding quark and antiquark to be
annihilated into a lepton pair, has no transverse momentum.
Therefore, the final state has zero transverse momentum, too.

The diagrams contributing to the Drell-Yan process in addition to
the LO parton model mechanism (a quark from one hadron annihilates
with an antiquark from another hadron into a virtual photon) are
divided into two classes: those of higher orders of $\alpha _S$
and those of higher twists. Calculating next-to-leading order
contributions and refitting the parton distributions accordingly,
one can reduce the discrepancy with the data on the
double-differential Drell-Yan cross section~\cite{NLOK,BFKL}.
Moreover, the higher twists are vanishing in the limit of infinite
energy. However, these power suppressed contributions can be large
at realistic energies. We show in the present paper that the
contribution of higher twists is essential for a proper
description of the data on the triple-differential Drell-Yan cross
section and propose a phenomenological model suitable to calculate
these effects. We note that recently also H.~Shimizu~{\it et
al}~\cite{vogelsang} have pointed out that at lower energies the
dilepton cross section provides information on the
non-perturbative dynamics.

\begin{figure}
\begin{center}
 \resizebox{0.65\textwidth}{!}{%
   \includegraphics{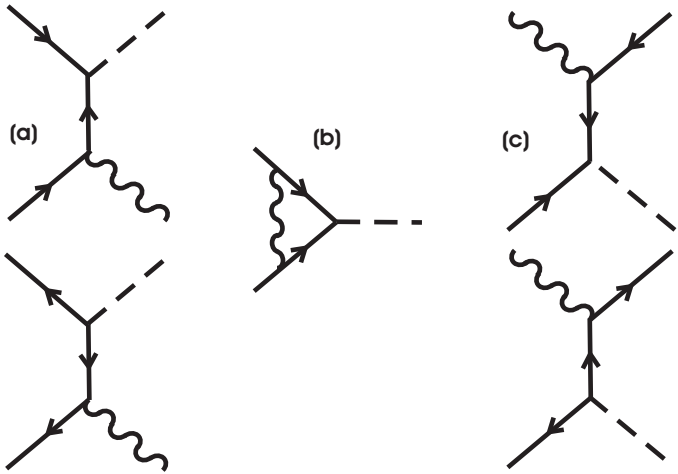}
 }
\caption{$O(\alpha_S)$ contributions to the Drell-Yan process: (a)
gluon Bremsstrahlung, (b) vertex correction, (c) gluon Compton
scattering. Virtual photons (dashed lines) split into lepton
pairs, waved lines denote gluons, arrows denote quarks. In each
diagram, time runs from left to right.} \label{NLO}
\end{center}
\end{figure}

The lepton pairs can gain non-vanishing $p_T$ due to two possible
mechanisms
\begin{itemize}
\item processes of next-to-leading order~(NLO) in $\alpha_S$,
\item quark transverse motion and off-shellness.
\end{itemize}
In the former case, the dilepton pair recoils against an
additional jet in the final state. An example is given by the
production of a gluon besides the virtual photon (see
Fig.~\mbox{\ref{NLO}a})
\be \label{qqg} \bar q q \to l^+l^-g. \ee
Al\-ter\-native\-ly, fi\-nal trans\-verse mo\-ment\-um of the
lepton pair can be caused by a non-vanishing transverse momentum
of the initial state, {\it i.e.} by the non-collinearity of quarks
inside the colliding hadrons. In this case, the recoil transverse
momentum is carried by hadron remnants, formed by the ``spectator"
partons. In addition, the quark and gluon off-shellness can have a
large effect for some observables, as has been recently shown
in~\cite{collins.jung} and~\cite{paper2}. The parton off-shellness
arises due to interaction between the partons in one hadron and
thus constitutes a higher twist effect.

In the present paper, we compare the Drell-Yan process cross
section and the transverse momentum $p_T$ of Drell-Yan pairs in
both aforesaid approaches.
We calculate in collinear pQCD the mean $\langle p_T ^2 \rangle
_{pert}$, which is the part of the lepton $p_T^2$ generated by the
next-to-leading order process (\ref{qqg}). This
allows us to study the evolution of $\langle p_T ^2 \rangle
_{pert}$ with $s$ and $M$. However, it turns out that the
magnitude of the experimentally measured $p_T$ cannot be described
by NLO alone. On the other hand, the experimental data are
reproduced much better by a model, taking into account both the
intrinsic transverse momentum and the off-shellness of quarks in
the proton, as will be shown in section~\ref{sect3}. In this model
there is no need for a $K$-factor. Instead, we describe both shape
and magnitude of double as well as triple differential cross
sections by extracting a quark-off-shellness from the data, which
allows a physical interpretation. We expand the cross section in
this model in powers of $1/s$ at $s\to \infty$ in order to relate
the phenomenological result to higher twist corrections in
section~\ref{sect4}. We conclude in section~\ref{conclusions}.

\begin{figure*}
\begin{center}
\subfigure
{
    \resizebox{0.48\textwidth}{!}{%
       \includegraphics{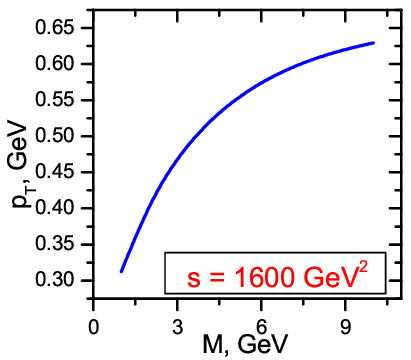}
    }
\label{c} } \hspace{-0.5 cm}
\subfigure
{
   \resizebox{0.48\textwidth}{!}{%
        \includegraphics{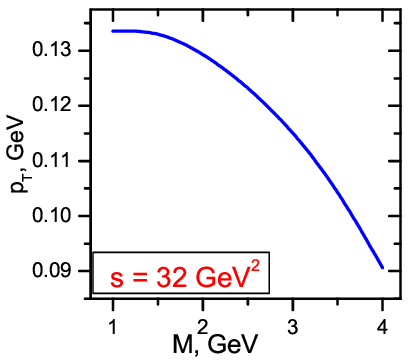}
   } \label{d}
} \caption{Perturbatively generated $ p_T \equiv \sqrt{ \langle
\vec p \, ^2 _T \rangle _{pert} } $ as a function of the invariant
mass M of the dilepton pair.} \label{e}
\end{center}
\end{figure*}


\section{Effects of non-collinearity and off-shellness of quarks in the proton}
\label{sect3}

The transverse momentum distribution of Drell-Yan dileptons $d
\sigma/d M^2 d x_F d p_T^2$ in the NLO perturbative QCD is
singular at $p_T=0$
~\cite{Altarelli,NLO_vs_data}. Neither can the average $p_T$ of
Drell-Yan pairs be reproduced in NLO: Using the method introduced
in~\cite{Altarelli}, we calculate $\langle \vec p_T^2 \rangle$ of
dileptons in NLO perturbative QCD. The results are presented in
Fig.~\ref{e}. The $p_T$ width calculated in this way is around
$0.6$~GeV at $s=1600$~GeV$^2$ and $M\approx 7$~GeV (see
Fig.~\ref{c}). This value is about a factor of 2 smaller than the
width of the $p_T$ distribution measured by the Fermilab
experiment E866~\cite{E866}. We have to conclude that NLO pQCD is
insufficient to describe present data on $p_T$ of Drell-Yan pairs.

In this context, we note an interesting observation: while the
perturbative $\langle p_T^2 \rangle _{pert}$ of Drell-Yan pairs
increases with increasing $M$ at high center of mass energy
$\sqrt{s}$ (see Fig.~\ref{c}), it decreases with $M$ at low $s$
(for instance, at $s=32$~GeV$^2$ relevant for the future PANDA
experiment, Fig.~\ref{d}) due to phase-space limitations.

A natural approach to generate additional $p_T$ is to take into
account the primordial transverse momentum of quarks in the
proton. The primordial quark transverse momentum is a
non-perturbative effect
and, from the uncertainty principle, averages at $\gtrsim
200$~MeV. On the other hand, we will show that the higher twist
effect of the intrinsic $k_T$ on the Drell-Yan cross section is of
the same order as the effect of non-vanishing quark off-shellness
in proton, which is caused by the interaction of partons of one
hadron in the initial state. Therefore, both the intrinsic $k_T$
and quark off-shellness have to be taken into account for the sake
of consistency.

The data on triple differential cross section also favor a model
taking into account both non-collinearity and off-shellness of
quarks. Such a model was proposed and tested against the E866
data~\cite{E866} in earlier publications~\cite{paper3,paper2}. The
method utilizes a phenomenological transverse momentum
distribution and an off-shellness distribution
of quarks in the proton.

This approach is based on the factorization assumption and on a
second assumption that the soft part of the cross section can be
approximated by a product of functions of the quark $k^+$
momentum, the transverse momentum, and the virtuality $m^2\equiv
k^+k^- - \vec k \, ^2_T$. The part dependent on $k^+$ and $k_{T}$
is
\be \label{unintPDF} g _q (M ^2, \xi, \vec k_T) = g(\vec k_T)
q(\xi,M^2), \ee
where $q(\xi,M^2)$ denotes parton distribution of quark flavor $q$
at momentum fraction $\xi$ and scale $M^2$. We use a Gaussian
distribution of quark intrinsic transverse momentum $k_T$
\be
\label{D} g(\vec{k}_T) = \frac{1}{4 \pi D^2} \exp \left( -\frac{
\vec{k} \,_{T}^2}{4 D^2} \right) , \ee
the mean squared partonic intrinsic transverse momentum being
$\langle \vec{k}\,_T  ^2 \rangle = 4 D ^2$. We note that both the
factorization ansatz (\ref{unintPDF}) and the Gaussian
parametrization (\ref{D}) are model assumptions, which are open
for improvement~\cite{Close.Halzen.Scott}, even though they are
widely used at present~\cite{Efremov}.

The quark virtuality distribution cannot be calculated from first
principles. Using analogy to many body theory, we parametrize the
quark virtuality distribution as a Breit-Wigner with width
$\Gamma$
\be
\label{BW} \mbox{A}(m) =
\frac{1}{\pi}\frac{\Gamma}{m^2+\frac{1}{4}\Gamma^2} . \ee
The exact off-shell kinematics as well as the off-shell and
non-collinear sub-process cross section (at LO in $\alpha_S$) are
used.

The cross section of the process $'hadron\ A'+'hadron\  B'\to
l^+l^+X$ in this approach is~\cite{paper2}
\bea \label{hadronDY} \frac{d ^3 \sigma }{dM^2dx_Fdp_T^2} \! & \!
= \! & \! \sum _q \! \int \! d \vec{k}_{1\perp} \! \int \! d
\vec{k}_{2\perp} \! \int _0 ^\infty \! d m_1 \! \int _0 ^\infty \!
d m_2 \! \int _0 ^1 \! d \xi_1 \! \int _0 ^1 \! d \xi_2 \mbox{A}
(m_1) \mbox{A} (m_2) \nn \! &\! \!& \! \times
g^A_q(M^2,\xi_1,\vec{k}_{1\perp}) \bar
g^B_q(M^2,\xi_2,\vec{k}_{2\perp}) \frac{d ^3 \hat \sigma _q
(m_1,m_2,\vec k_{1\perp},\vec k_{2\perp}) }{dM^2dx_Fdp_T^2},\ \eea
in which not only the three-dimensional (longitudinal and
transverse) motion of partons, but also the virtualities of the
active quark and anti-quark are explicitly taken into account by
means of a phenomenological double-unintegrated parton density.

In (\ref{hadronDY}), the following off-shell partonic cross
section is used
\bea \label{partonDY} \frac{d \hat \sigma }{dM^2dx_Fdp_T^2} \! \!
& \! \!   = \! \!   &  \! \!  \kappa' \left[
     2 M^4 - M^2 \left( m_1^2 - 6 m_1 m_2 + m_2^2 \right) - \left( m_1^2-m_2^2 \right) ^2
\right] \nn
&&
\! \! \! \! \! \! \! \! \! \! \! \! \! \! \! \! \! \! \! \! \! \!
\! \! \! \! \! \! \! \! \! \!
\times \delta \left( \! \! M^2
        -m_1^2-m_2^2 - \xi_1 \xi_2 P_1 ^- P_2 ^+
        - \frac{\left( m_1^2 +\vec{k}_{1\perp}^2 \right) \!
                \left( m_2^2 +\vec{k}_{2\perp}^2 \right)}{\xi_1\xi_2 P_1^- P_2^+}
        + 2 \vec{k}_{1\perp} \cdot \vec{k}_{2\perp}
\! \right) \! \nn
&&
\! \! \! \! \! \! \! \! \! \! \! \! \! \! \! \! \! \! \! \! \! \!
\! \! \! \! \! \! \! \! \! \!
\times \delta\left( \! x_F - \frac{\sqrt{s}}{s-M^2} \left\{ \xi_2
P_2 ^+  - \xi_1 P_1 ^- + \frac{\left( m_1^2 +\vec{k}_{1\perp} ^2
\right)}{\xi_1 P_1 ^-} - \frac{\left( m_2^2 +\vec{k}_{2\perp} ^2
\right)}{\xi_2 P_2 ^+} \right\}  \! \right) \! \!  \nn
&&
\! \! \! \! \! \! \! \! \! \! \! \! \! \! \! \! \! \! \! \! \! \!
\! \! \! \! \! \! \! \! \! \!
\times \delta \left( \left( \vec{k}_{1\perp}+\vec{k}_{2\perp}
\right)^2 -p_T^2 \right), \eea
with
\be
\label{DY5} \kappa' = \frac{2 \pi \alpha^2 e_q^2 
}{3 M^4 N_c 8 \sqrt{(k_1\cdot k_2)^2-m_1^2 m_2^2}}. \ee
In (\ref{partonDY}),
\be (k_1 \cdot k_2) = \xi_1 \xi_2 P_1^- P_2^+ + \frac{(m_1^2+\vec
k \, ^2_{1 \perp})(m_2^2+\vec k \, ^2_{2 \perp})}{ \xi_1 \xi_2
P_1^- P_2^+} - \vec k_{1 \perp} \vec k_{2 \perp}, \ee
\be
\label{P_in_CMS} \left( P_1^\mp \right)^2 = \left( P_2^\pm
\right)^2
=
\frac{s}{2}-M_N^2\pm \sqrt{ \left( \frac{s}{2} \right)^2-M_N^2 s}
\ee
in the hadron center of mass system\footnote{
The formula (\ref{partonDY}) is a corrected version of the
erroneous formula (39) in~\cite{paper2} rewritten in the Lorentz
invariant form. Numerically, the difference between the two
expressions is less than 1\% in the kinematics studied
in~\cite{paper2} but might be large at $x_F \to 1$ and $p_T \gg
M$.
}.

In the scaling limit ($s\to \infty$, $s/M^2=$const), the spectral
functions effectively drop out due to normalization ({\it cf.}
discussion later), and the hadronic cross section (\ref{hadronDY})
goes to
\be
\label{LT} \frac{ d ^3 \sigma}{d M^2 d x_F  d p_T ^2 } = \sum _q
\Phi _q (x_1, x_2) \left( \frac{d ^2 \hat \sigma _q}{ d M^2 d x_F
} \right)_{LO} \frac{1}{8 D^2 } \exp \left( - \frac{ p_T ^2}{ 8
D^2} \right), \ee
where
\be \label{Phi} \Phi _q (x_1, x_2) \equiv q ^A (x_1) \bar q ^B
(x_2) + \bar q ^A (x_1) q ^B (x_2), \ee
\be
\left( \frac{d ^2 \hat \sigma _q}{ d M^2 d x_F } \right)_{LO} =
\frac{4 \pi \alpha^2 e_q^2 }{9 M^4 } \frac{x_1 x_2 }{x_1 + x_2}
(1-x_1 x_2), \ee
and the parton momentum fractions ($x_1$, $x_2$) are defined via
\bea \label{x1x2} M^2 & = & x_1 x_2 s; \\ x_F & = &
(x_2-x_1)/(1-x_1 x_2). \eea
Note that the $x_1 x_2$ in the denominator of the $x_F$ definition
in (\ref{x1x2}) is sometimes omitted in textbooks, where an
approximate definition $x_F \approx 2 p_z/ \sqrt{s}$ is used
instead of (\ref{xf}).

\begin{figure}
\begin{center}
        \resizebox{0.75\textwidth}{!}{%
        \includegraphics{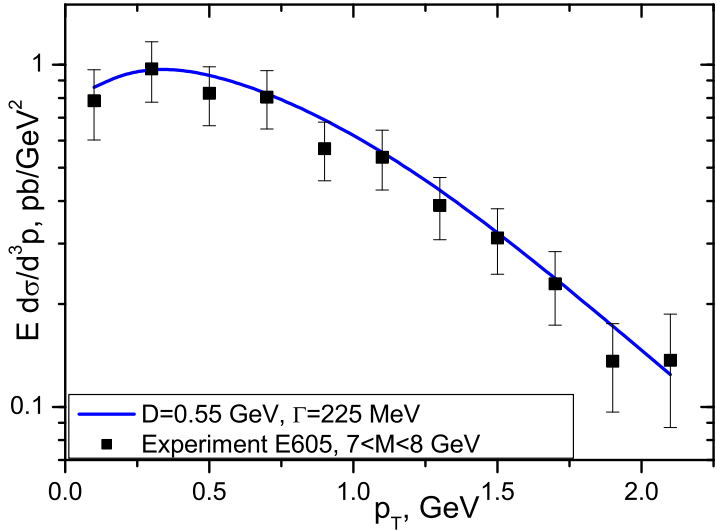}
        }
\caption{Prediction for the $p_T$ spectrum of Drell-Yan dileptons
in our model at $\sqrt{s}=40$~GeV and $x_F=0.1$ as compared to the
data of experiment E605~\cite{E605}.} \label{E605}
\end{center}
\end{figure}

In~\cite{paper2}, our model was compared to the data on the triple
differential Drell-Yan cross section $d ^3 \sigma / d M^2 d x_F d
p_T $ from experiment E866~\cite{E866} at Fermilab in $p p$
collisions at 800~GeV incident energy. Both the slope and
magnitude of the $p_T$ distribution of the Drell-Yan pairs were
described well without the need for a $K$-factor. In particular,
the experimentally measured $\langle p_T^2 \rangle$ is reproduced
in this model by fitting the model parameter $D$ (the dispersion
of the quark intrinsic transverse momentum). At $s=1600$ GeV$^2$,
we obtained $D=0.5\pm 0.18$~GeV. On the other hand, the detailed
shape of the distribution turned out to be sensitive to the
off-shellness, giving $\Gamma\!=\!50\!-\!300$~MeV (depending on
the mass bin) for this particular experiment.

The distribution of the transverse momentum of lepton pairs
produced in the Drell-Yan process off {\em nuclei} $p A \to
l^+l^-X$ also can be reproduced within this model. For example, in
Fig.~\ref{E605} the calculation for the transverse momentum
spectrum of Drell-Yan dileptons of our model is compared to the
data of the experiment E605~\cite{E605} on $p\  C u$ collisions at
$\sqrt{s}=38.8$~GeV, $x_F=0.1$. The cross section plotted in
Fig.~\ref{E605} is
\be
E \frac{d^3\sigma}{d ^3 p} \equiv  \frac{2 E}{\pi
\sqrt{s}}\frac{d\sigma}{d x_F d p_T ^2} =\frac{2 E}{\pi \sqrt{s}}
\int \limits _{\mbox{\small bin}} \!
                                \frac{d\sigma}{dM^2dx_Fdp_T^2} \, dM^2 .
\label{triple} \ee
where $E$ is given by
\be
E\equiv \sqrt{M^2+p_T^2+x_F^2 (s-M^2)^2/(4 s)}. \ee
The model parameters $D$, $\Gamma$ used in the calculations were
fitted to data on $p p \to l^+l^-X$ in~\cite{paper2} and no
readjustment was done for the $p A$ case.

\begin{figure}
\begin{center}
        \resizebox{0.75\textwidth}{!}{%
        \includegraphics{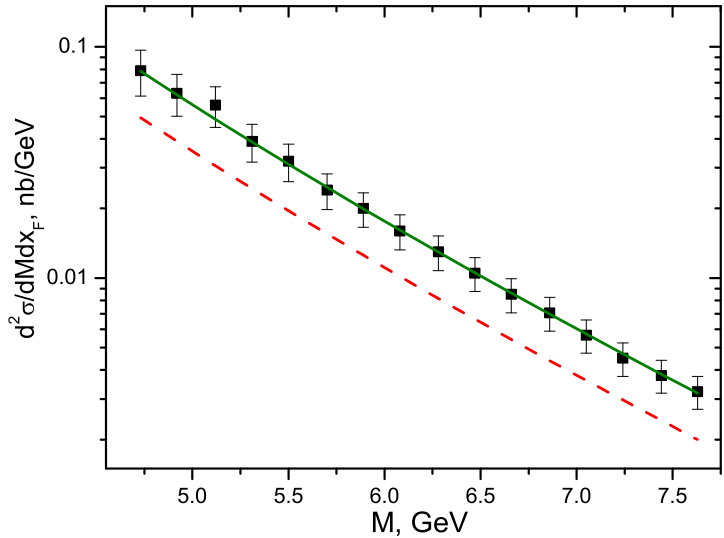}
        }
\caption{Result of LO collinear QCD for a double differential
Drell-Yan cross section (dashed line) at $\sqrt{s}=20$~GeV and
$x_F=0.1$ as compared to experiment E439~\cite{E439}. Solid line
is the LO result scaled up with a factor $K=1.6$.} \label{LO}
\end{center}
\end{figure}

In the present paper, we want to make a consistent comparison with
NLO results. However, the triple differential cross section $E
d^3\sigma/d ^3 p$, plotted in Fig.~\ref{E605}, cannot be compared
directly to the NLO result, because the latter is singular in
every fixed order of pQCD~\cite{diverg,BFKL}. We, therefore, apply
the described model now to the double differential cross section
$d ^2 \sigma / d M d x_F$. The $K$-factor, which is needed to
increase the magnitude of the LO prediction for the double
differential Drell-Yan process cross section so that it agrees
with the data, can be decreased from 2 to 1.1 by taking into
account NLO processes~\cite{NLOK}. In order to determine, what
part of this LO $K$-factor can be accounted for by the model with
intrinsic $k_T$ and off-shellness of quarks, we compare data to
the triple differential cross section~(\ref{hadronDY}) integrated
over $p_T$.

In Fig.~\ref{LO}, the Drell-Yan process cross section $d ^2 \sigma
/ d M d x_F$ predicted at leading order of perturbative QCD
(dashed line) is compared to the data of the Fermilab experiment
E439~\cite{E439} on $p W$ collision at 400~GeV incident energy, at
$x_F = 0.1$. The LO prediction lies below the data. The solid line
shows the LO curve scaled up with a factor $K=1.6$. The $K$-factor
depends somewhat on the parametrization of parton distributions
used. We use here the parametrization~\cite{grv}. If one assumes a
larger contribution of sea quarks, the $K$-factor needed to
describe the data is lower.

\begin{figure*}
\begin{center}
        \resizebox{0.8\textwidth}{!}{%
        \includegraphics{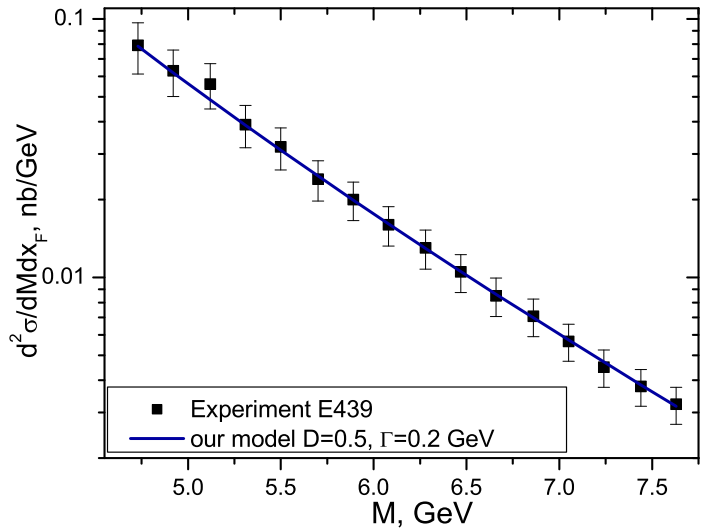}
        }
\caption{ Drell-Yan cross section in our model at $D=500$~MeV,
$\Gamma=200$~MeV compared to the data of experiment
E439~\cite{E439}. $K=1$.} \label{off}
\end{center}
\end{figure*}

In order to calculate the double differential Drell-Yan process
cross section in our model, we first find the triple differential
cross section from (\ref{hadronDY}). The double differential cross
section is obtained using
\be
\frac{d ^2 \sigma }{ d M^2 d x_F} \equiv \int _0 ^{(p_T^2)_{max}}
\frac{d ^3 \sigma}{ d M^2 dx_F dp_T^2} dp_T^2. \label{int} \ee
Note that the maximum transverse momentum of the Drell-Yan pair
$(p_T^2)_{max}$ is fixed by kinematics
\be (p_T^2)_{max} = \frac{(s+M^2- M_R ^2)^2}{4 s} - x_F^2
\frac{(s-M^2)^2}{4 s} - M^2, \label{pTmax} \ee
where $M_R^2$ is the minimal invariant mass of the undetected
remnant.


The data are reproduced well (see Fig.~\ref{off}) with $K=1$. We
conclude that in the experimentally relevant region the $K$-factor
of the double differential Drell-Yan cross section can be
explained by two alternative scenarios: either as an effect of
higher orders of perturbative QCD as shown in~\cite{NLOK,BFKL} or
as an effect of non-collinearity and off-shellness of quarks in
our phenomenological approach. The experimental cross section
magnitude can be reproduced in NLO calculations by fitting the
renormalization scale or  in our model by fitting the parameters
$D$ and $\Gamma$. The latter explanation of the $K$-factor has the
advantage that it can explain also the triple differential cross
section.

In the following, we additionally study the relative importance of
quark off-shellness
and quark intrinsic transverse motion by comparing our result to
that of the intrinsic-$k_T$ approach. The intrinsic-$k_T$
approach~\cite{all_kt} is a limiting case of our model at $\Gamma
\to 0$. The factorization assumption in this case gives
\be
\frac{ d ^4 \sigma }{ d M^2 d x_F d \vec p_T }= g (\vec k_{T1})
\otimes g (\vec k_{T2}) \otimes \frac{d ^2 \hat \sigma (\vec
k_{T1}, \vec k_{T2}) }{ d M^2 d x_F} \delta (\vec p_T - \vec
k_{T1} - \vec k_{T2}). \label{gfact} \ee

\begin{figure}
\begin{center}
        \resizebox{0.75\textwidth}{!}{%
        \includegraphics{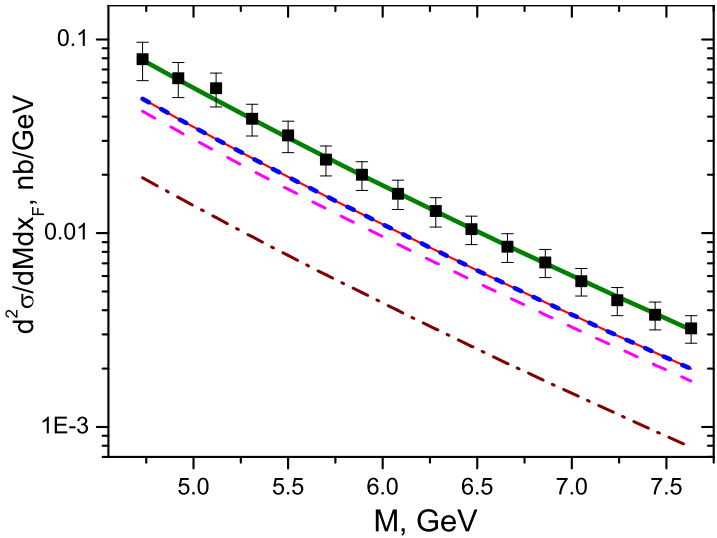}
        }
\caption{Drell-Yan cross section in a simplified intrinsic-$k_T$
approach at $D=50$~MeV (short dash), $D=250$~MeV (dash),
$D=500$~MeV (dash-dot) compared to LO pQCD (thin solid line) and
the data of experiment E439~\cite{E439}. The solid line gives the
theoretical curves multiplied by $D$-dependent $K$-factors fitted
to the data. Everywhere, $\Gamma=0$.} \label{KT}
\end{center}
\end{figure}

The formula is often simplified by neglecting the dependence of
$\hat \sigma $ on $\vec k_{T1}$ and $\vec k_{T2}$, for example
in~\cite{Alesio.Murgia.D_vs_S} and in PYTHIA~\cite{PYTHIA}. In
this case, the $p_T$ spectrum of Drell-Yan pairs $d ^3 \sigma / d
M^2 d x_F d p_T^2$ is also simply a Gaussian in $p_T^2$. The cross
section (\ref{gfact}) has to be integrated over the azimuthal
angle of the lepton pair and over $p_T^2$ according
to~(\ref{int}). Because of the finite integration interval
in~(\ref{int}), we do not recover the normalization of the
$k_T$-distribution~(\ref{D}), but obtain a suppression that
increases with $D$.

The double differential Drell-Yan process cross section in the
intrinsic $k_T$ approach with collinear sub-process cross section
at three values of $D$ is compared to the LO of pQCD and the data
of the experiment E439 in Fig.~\ref{KT}. The magnitude of the
measured cross section cannot be reproduced in this leading order
intrinsic $k_T$ approach, in which the dependence of the partonic
$d \hat \sigma$ on $\vec k _{1T}$ and $\vec k _{2T}$ are
neglected. Just as in LO of pQCD, scaling with an overall
$K$-factor ranging from $1.6$ to $4$ is necessary to describe the
data. It is apparent from Fig.~\ref{KT} that the $K$-factor
extracted from the data is $D$-dependent. Therefore, this scaling
factor should be understood as a phenomenological parameter and
not as a measure of higher order corrections.

\begin{figure}
\begin{center}
        \resizebox{0.75\textwidth}{!}{%
        \includegraphics{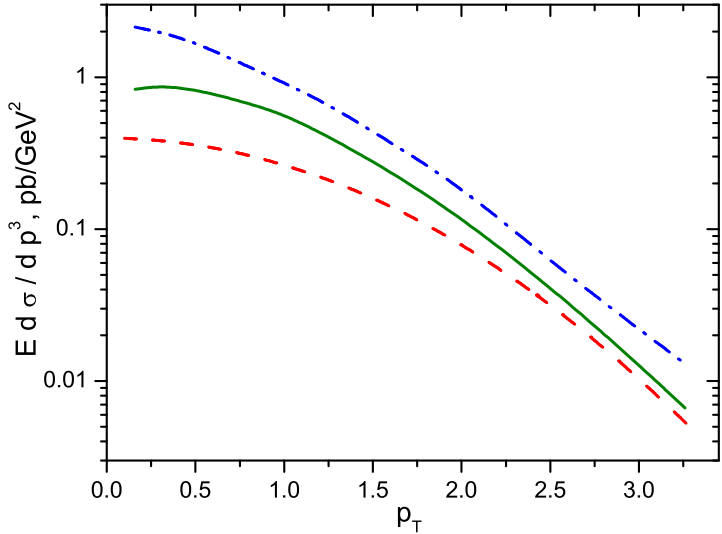}
        }
\caption{Drell-Yan cross section in three approaches: simplified
intrinsic-$k_T$ with collinear sub-process cross section (dash),
full intrinsic-$k_T$ (dash-dot), our model at $\Gamma=225$~MeV
(solid). Everywhere, $D=550$~MeV, $\sqrt{s}=40$~GeV,
$M\approx7.5$~GeV. } \label{3models}
\end{center}
\end{figure}

Summarizing, there are three phenomenological approaches to the
Drell-Yan process beyond LO pQCD:
\begin{enumerate}
\item our model, accounting for both intrinsic transverse
momentum ($D\ne 0$) and off-shellness ($\Gamma \ne 0$) of quarks;
\item intrinsic-$k_T$ approach ($D\ne 0$), which is the limiting
case of our model at $\Gamma=0$;
\item simplified intrinsic-$k_T$ approach ($\Gamma =0$),
in which the primordial transverse momentum is not zero ($D\ne
0$), but the non-collinearity of the $\bar q q \to l^+l^-$
sub-process cross section $d\hat \sigma$, {\it i.e.} its
dependence on $\vec k_1$ and $\vec k_2$, is neglected.
\end{enumerate}
We compare the effects of primordial $k_T$, non-collinearity of $d
\hat \sigma$ and quark off-shellness by plotting the triple
differential Drell-Yan cross section calculated in the three
aforesaid phenomenological approaches in Fig.~\ref{3models}. The
simplified intrinsic-$k_T$ approach gives a Gaussian for the
$p_T$-distribution (dash line). As we will show in the next
section, the approximation of $\Gamma=0$ and collinear $d \hat
\sigma$ is equivalent to restricting oneself to the leading order
in the twist expansion, that is, in the case of the unpolarized
Drell-Yan process, the expansion in powers of $1/M$. In
Fig.~\ref{3models}, the importance of higher twist corrections in
the Drell-Yan process is illustrated by the difference between the
solid and dash lines.

The part of higher-twist effects incorporated in the full
intrinsic-$k_T$ approach changes the distribution considerably
({\it cf.} the dash and dash-dot curves in Fig.~\ref{3models}). On
the other hand, additional higher twist effects, modelled by quark
off-shellness and given by the difference between the dash-dot and
solid curves, are of the same order. We conclude that higher
twists in the Drell-Yan process can be large and that we have to
take into account both non-collinearity and off-shellness of
quarks in order to model them.



\section{Twist nature of the phenomenological corrections}
\label{sect4}

\begin{figure}
\begin{center}
        \resizebox{0.8\textwidth}{!}{%
        \includegraphics{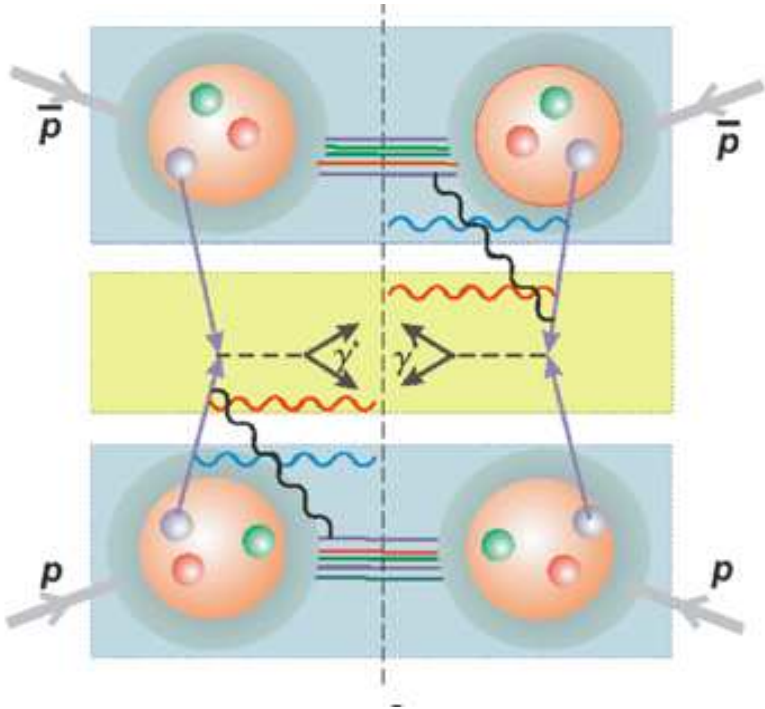}
        }
\caption{Example of gluon radiation diagrams generating intrinsic
$k_T$ and virtuality of quarks in the proton as probed in the
Drell-Yan process.} \label{DiagramDY}
\end{center}
\end{figure}

In the previous section we have shown that the double differential
Drell-Yan cross section is reproduced by a model accounting for
intrinsic $k_T$ and off-shellness of quarks without a need for a
$K$-factor. In addition, the $p_T$ distribution of the Drell-Yan
pairs can be explained in our model~\cite{paper2}, but not in NLO
of pQCD~\cite{diverg,NLO_vs_data}. Therefore, the effects of quark
off-shellness and intrinsic $k_T$ do not arise solely from the
diagrams of NLO pQCD. Instead, we will show that they parametrize
higher twist processes. Some of the diagrams that contribute to
the Drell-Yan cross section at higher twist are shown in
Fig.~\ref{DiagramDY}. Gluon radiation in the initial state and
gluon exchange between the active parton and spectators generate
intrinsic $k_T$ and virtuality of quarks in the proton in the
Drell-Yan process. Some of these processes (for example, the gluon
exchanges that connect factorized regions - the sub-process and a
soft matrix element) are suppressed by powers of $s$ in the
scaling limit. However, the power-suppressed corrections give a
sizable contribution to the transverse momentum spectrum of
Drell-Yan pairs at finite $s$ accessible in modern experiments.

In this section, we investigate the relationship of NLO and higher
twist corrections to those calculated in our phenomenological
approach by comparing their behaviour in the Drell-Yan scaling
limit, in which $s \to \infty$ and $M^2 \to \infty$ with
$\tau=s/M^2$ finite. The NLO corrections are proportional to
$\alpha _S$; therefore we expect only a logarithmic dependence of
these corrections on the hard scale $s \sim M^2$. On the other
hand, higher twist contributions are suppressed in powers of $s$
in the scaling limit. In order to determine whether the effects of
quark virtuality and intrinsic $k_T$ are leading twist, we study
the behaviour of the Drell-Yan cross section calculated in our
model in the scaling limit.

For comparison, note that the lowest order triple differential
cross section is given by
\be
\left ( \frac{ d ^3 \sigma}{d M^2 d x_F  d p_T ^2 } \right) _{LO}
= \sum _q \Phi _q (x_1, x_2) \left( \frac{d ^2 \hat \sigma _q}{ d
M^2 d x_F } \right)_{LO} \delta \left( p_T^2 \right). \ee
From (\ref{LT}) one observes that the corrections to the $p_T$
distribution of the Drell-Yan pairs due to quark non-collinearity
in the proton are not suppressed in the limit $s \to \infty$. The
model taking into account the intrinsic $k_T$ of partons therefore
parametrizes some of the leading twist effects. This part of the
model effects, {i.e.} the Gaussian smearing of $p_T$, is related
to contributions of the higher order diagrams of the perturbative
QCD series as is shown for deep inelastic scattering at small
Bjorken $x$ in~\cite{BFKL}.

On the other hand, the corrections to the LO cross section
generated by the $k_T$ dependence of the sub-process cross section
$d \hat \sigma$ are suppressed by powers of the hard scale $s$.
Therefore, they represent part of the higher twist effects. To
study this in more detail, we expand the cross section
(\ref{hadronDY}) in a series in $1/s$ around $s=\infty$, keeping
this time not only the leading term, as it has been done in
(\ref{LT}), but all the terms that are suppressed by less than
$s^{3/2}$. We analyze the cross section at the specific value of
$x_F=0$ to make the formulas less bulky.

We start from the general formula (\ref{hadronDY}). First, we
expand the integrand of (\ref{hadronDY}) in $1/s$. For this
purpose not only $d \hat \sigma$ of (\ref{partonDY}) has to be
evaluated at $s \to \infty$, but also the combination of parton
distributions (\ref{unintPDF}) that enters (\ref{hadronDY}) has to
be Taylor expanded around $(\xi_1 = \sqrt{ \tau}, \xi_2 =
\sqrt{\tau})$. The arguments of parton distributions $\xi_1$ and
$\xi_2$ are fixed after integrating out the $\delta$-functions in
(\ref{partonDY}). As a result, the probed parton light cone
momentum fractions depend on quark intrinsic transverse momentum
and off-shellness. After integrating (\ref{hadronDY}) over
$\xi_1$, $\xi_2$ and angles, the quantity $\Phi_q(\tilde \xi_1,
\tilde \xi_2)$ enters the hadronic cross section formula. Here,
$\tilde \xi_1$ and $\tilde \xi_2$ are
\bea \tilde \xi_1 & = & \sqrt{\tau} \left(1 + \frac{p_T^2/2 -
m_2^2 - k_2^2}{ \sqrt{ \tau} s} + O\left( \frac{1}{s^2} \right)
\right),
\\ \tilde \xi_2 & = & \sqrt{\tau} \left( 1 + \frac{p_T^2/2
-m_1^2-k_1^2}{\sqrt{\tau} s} + O\left( \frac{1}{s^2}
\right)\right). \eea
Keeping the first two orders in the Taylor expansion of
$\Phi_q(\tilde \xi_1, \tilde \xi_2)$ and in the \mbox{$1/(\tau
s)$-expansion} of $d \hat \sigma$, we obtain:
\begin{eqnarray} \label{big}
\! \! \! & \! \! \! \! \! \! & \! \! \! \left. \frac{ d\sigma ^3
_q }{d M^2 d x_F d p_T^2} \right| _{x_F=0} \! \! \! \! = \! \!
 \frac{\alpha ^2 e_q ^2 (1 - \tau)}{ 8 \pi D^4 6 \tau ^2 \sqrt{\tau} s ^3 }
 \int _0 ^\infty \!  dk_2^2 \! \int _{
(k_1^2)_{min}} ^{(k_1^2)_{max}} \! dk_1^2 \! \int _0
^{(m_2)_{max}} \! dm_2 \! \int _0 ^{(m_1)_{max}} \! \! \!\! \! \!
dm_1
 \nn
 & \! \! \! \! &
\phantom{\frac{ d\sigma ^3 _q }{d M^2 d x_F d p^2}} \! \times \!
\frac{  A(m_1) A(m_2) \exp \! \left( - \frac{k_1^2+k_2^2}{4 D^2}
\right) }{\sqrt{ k_1^2 k_2^2 - \frac{1}{4}(p_T^2-k_1^2 -k_2^2)^2
}} \left[ G_1 ^q (\tau) \frac{\sqrt{\tau}}{8} \left( \!
\frac{p_T^2}{2}-m_2^2-k_2^2\right) \right. \nn
 & \! \! \! \! &
\phantom{\frac{ d\sigma ^3 _q }{d M^2 d x_F d p^2}} \! + \!
G_2 ^q (\tau) \frac{\sqrt{\tau}}{8} \left( \!
\frac{p_T^2}{2}-m_1^2-k_1^2\right) \nn
 & \! \! \! \! & \left. \! \phantom{\frac{ d\sigma ^3 _q }{d M^2
d x_F d p^2}} \! + T^q (\tau) \left( \frac{\tau s}{8} +
\frac{p_T^2}{16}
+ F(m_1,m_2) \right)\! + O \!
\left(\frac{1}{s}\right) \! \right] \!,
\end{eqnarray}
where
\bea (k_1^2)_{min} & \equiv & (p_T-k_2)^2; \\
(k_1^2)_{max} & \equiv & (p_T+k_2)^2; \\
(m_1)_{max} & \equiv & \sqrt{\tau s + p_T^2/2 - k_1^2} ;
\label{mlim1}
\\
(m_2)_{max} & \equiv & \sqrt{\tau s + p_T^2/2 - k_2^2};
\label{mlim2}
\\
F(m_1,m_2) & \equiv & \frac{1}{\tau} \left( 2(m_1^2+m_2^2) (\tau
/8 + \tilde\xi_1^2 /6 - \tilde\xi_1 \sqrt{\tau}/6 ) + (m_1+m_2)^2
\tau / 6 \right. \nn
&&  + m_1^2 \tau/6 +(m_1^2-m_2^2)\tilde\xi_1 \sqrt{\tau} / 6 -
m_1^2 \tau \sqrt{\tau} / (6 \tilde\xi_1) \nn
&&\left. +\frac{\tau}{8} (m_1^2-m_2^2+k_1^2-k_2^2) \right); \\
T^q (\tau) & \equiv & \Phi ^q ( \sqrt{\tau} ,\sqrt{\tau}); \eea
and
\bea
G_1 ^q(\tau) & \equiv & \left. \frac{ \partial \Phi ^q ( x_1 ,x_2)
}{ \partial x_1} \right| _{(x_1=\sqrt{\tau},x_2=\sqrt{\tau})}, \\
G_2 ^q(\tau) & \equiv & \left. \frac{ \partial \Phi ^q ( x_1 ,x_2)
}{ \partial x_2} \right| _{(x_1=\sqrt{\tau},x_2=\sqrt{\tau})} \eea
are the derivatives of the parton distribution product around
$(\sqrt{\tau},\sqrt{\tau})$. Note that the term in square brackets
in eq.~(\ref{big}) is not symmetric in $k_1$ and $k_2$. This is
due to the different integration boundaries present in the $k_1$
and $k_2$ integrations. Several $\delta$-functions are evaluated
to get from (\ref{hadronDY}) to (\ref{big}). In the course of
these calculations, it turned out to be useful to treat $\vec k
_{1 \perp}$ and $\vec k _{2 \perp}$ differently.

To further investigate the dependence of the integral (\ref{big})
on $s$, we have to specify the quark spectral function. Indeed,
the integration variables $m_1$ and $m_2$ at $s \to \infty $ can
be arbitrarily big, as can be seen from (\ref{mlim1}) and
(\ref{mlim2}). Therefore, only after the integration over $m_i$
has been performed can we judge whether any off-shellness
generated term is sub-leading in $s$ and how much it is
suppressed. On the other hand, the integration over $m_i$ provides
additional terms $\sim k^2/s$, interconnecting the off-shellness
and intrinsic $k_T$ effects.

In the following, we perform the analytical integration of
(\ref{big}), assuming different functional forms for the spectral
function $A(m)$:
\begin{enumerate}
\item
a Dirac delta-function $\delta (m)$,
\item
a Breit-Wigner function (Lorentz distribution) with a constant
parameter $\Gamma$, see (\ref{BW}).
\end{enumerate}

In the former case, the model reduces to the intrinsic-$k_T$
approach. Integrations over $m_i$ drop out, while the remaining
integrals over $k_1^2$ and $k_2^2$ can be done
analytically via Bessel functions. As the result, one finds the
leading term (\ref{LT}) plus $1/(\tau s)$ suppressed
contributions.

Let us now consider the second, more general, case. The cross
section for $A(m)=\delta (m)$ is the limiting case of the formulas
given below for a Breit-Wigner distribution (\ref{BW}) at $\Gamma
= 0$. Inserting the spectral function (\ref{BW}) into (\ref{big}),
performing all the integrations and keeping only the first few
leading terms in $1/M$, we obtain (note that $M^2=\tau s \to
\infty$, as $s \to \infty$)
\bea \label{star}
 \left. \frac{ d ^3 \sigma _q }{dM^2dx_Fdp_T^2} \right| _{x_F=0} \! \! \!
 = \!
 \frac{1}{8 D^2 } \exp \left( - \frac{ \vec p \, _T
^2}{ 8 D^2} \right) \sum _q \left( \frac{d ^2 \hat \sigma _q}{ d
M^2 d x_F } \right)_{LO} \left[  T ^q (\tau) \right. \nn \left.
+ \left\{ 4 \, T ^q (\tau) - \sqrt{ \tau} \, ( G_1^q(\tau) +
G_2^q(\tau)) \right\} \frac{1}{\pi} \frac{\Gamma}{M} \right. \nn
\left.
+ \left\{ \sqrt{\tau} \left( G_1^q(\tau) + G_2^q(\tau) \right)
\left( \frac{p_T^2}{4} -2 D^2 \right) + \frac{8}{3} T^q (\tau)
\left( \frac{5 p_T^2}{4} + D^2 \right)
 \right\} \frac{1}{M^2} \right. \nn \left.
+ \  O\left( \frac{\Gamma}{M^3} \right) \right].
\eea
At leading twist, the  Gaussian distribution of $p_T$ (\ref{LT})
is recovered. However, it is modified by the higher twists,
suppressed in the limit $s\to \infty$, but substantial at finite
$s$ accessible in experiment. The term proportional to
$1/M=1/\sqrt{\tau s }$ is $p_T$-independent and leads to an
overall enhancement of the cross section, while the
$p_T$-dependent terms proportional to $1/M^2$ additionally modify
the shape of the $p_T$ distribution.

The contribution of the off-shellness of quarks to  (\ref{star})
is given by the summands proportional to $\Gamma$. It is
suppressed by powers of $M$ and vanishes in the intrinsic $k_T$
approach, in which $\Gamma=0$. Thus, the model, which additionally
accounts for quark off-shellness, parametrizes more higher twist
effects than the intrinsic-$k_T$ approach alone.

It is interesting that the effects due to the finite quark width
$\Gamma$ appear in the expansion at odd powers of $1/M$ in
contrast to those due to the intrinsic-$k_T$. The first
$\Gamma$-dependent correction is
proportional to
$1/M=1/\sqrt{\tau s}$.
%
%
Therefore, the corrections due to the virtuality of quarks seem to
have a non-analytical dependence on $s$ as $(\tau s)^{-1/2}$. In
order to preserve analyticity of the cross section we have to
assume that the quark spectral function width $\Gamma$ has a
particular scaling behavior at large hard scale of the probe
$M=\sqrt{\tau s}$:
\be
\Gamma (M) \sim \frac{1}{M},\mbox{ as } \  M \to \infty. \ee
Then, in (\ref{star}), the terms proportional to $\Gamma/M$ and
the terms proportional to $1/M^2$ together constitute the dominant
higher twist correction to the leading result (\ref{LT}) in the
scaling limit.

\begin{figure*}
\begin{center}
        \resizebox{0.75\textwidth}{!}{%
        \includegraphics{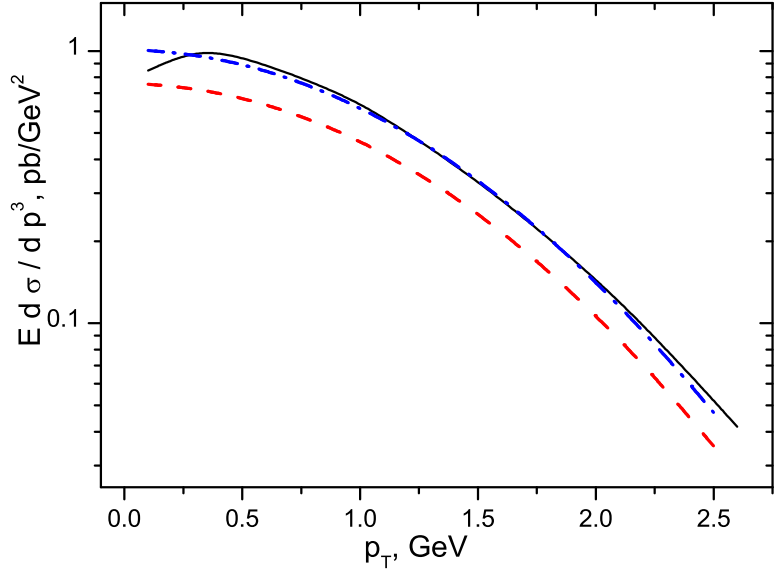}
        }
\caption{ Cross section of $pp\to l^+l^-X$ at $M \approx 7$~GeV,
$s=1600$~GeV$^2$ in our model (solid), in the leading order in
$1/M^2$ (dash), up to the next-to-leading order in $1/M^2$
expansion (dash-dot). $D=650$~MeV, $\Gamma=225$~MeV, $x_F=0$.}
\label{M7S1600}
\end{center}
\end{figure*}

\begin{figure*}
\begin{center}
        \resizebox{0.75\textwidth}{!}{%
        \includegraphics{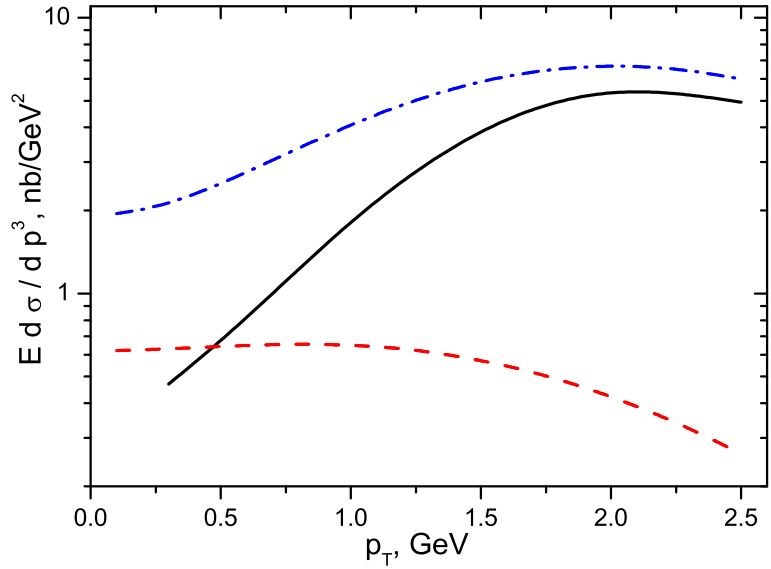}
        }
\caption{ Cross section of $p\bar p\to l^+l^-X$ at $M=1$~GeV and
$s=30.25$~GeV$^2$ in our model (solid), in the leading order in
$1/M^2$ (dash), up to the next-to-leading order in $1/M^2$
expansion (dash-dot). $D=600$~MeV, $\Gamma=250$~MeV, $x_F=0$.}
\label{M1S31}
\end{center}
\end{figure*}

We expect the formula (\ref{star}) to give a good approximation to
the Drell-Yan cross section (\ref{hadronDY}) at large finite $M$
and $s$. In order to illustrate this, we compare the result of the
exact calculations, {i.e.} the numerical integration of
(\ref{hadronDY}), to the leading twist approximation (\ref{LT})
and to the next-to-leading twist result (\ref{star}) in two
regimes:
\begin{itemize}
\item at $M \approx 7$~GeV and $s=1600$~GeV$^2$, see Fig.~\ref{M7S1600};
\item at $M=1$~GeV and $s=30.25$~GeV$^2$ relevant for FAIR~\cite{FAIR}, see Fig.~\ref{M1S31}.
\end{itemize}

As expected, the sum of leading and next-to-leading terms of the
power series (\ref{star}) reproduces the full calculations quite
well at $M$ as high as 7~GeV. The approximate cross section has
the same average magnitude and slope. Therefore, it is dominating
the $K$-factor type corrections to the leading twist cross
section. Only the bend of the cross section at low $p_T$, which is
seen in the full calculations and in the data (Fig.~\ref{E605}),
is not reproduced at the next-to-leading twist.

From the Fig.~\ref{M1S31}, one sees that our model predicts the
higher twist effects to be very large at low $M$ and $s$. The
discrepancy between approximate and exact Drell-Yan cross sections
is large in this regime, too, especially at low $p_T$. We conclude
that one needs to go beyond the next-to-leading twist at this low
$M$ and $s$. In this region, our model becomes indispensable,
because it effectively sums higher orders and higher twists.


\section{Conclusions}
\label{conclusions}

We have analyzed the double differential Drell-Yan cross section
$d ^2 \sigma / d M^2 d x_F $ and the $p_T$ distribution of the
Drell-Yan dileptons $d ^3 \sigma / d M^2 d x_F d p_T^2 $  in two
alternative approaches: collinear perturbative QCD at
next-to-leading order and a model, which makes use of
phenomenological distributions for $k_T$ and off-shellness of
quarks in the proton.

We find that the transverse momentum spectrum of the Drell-Yan
pairs at the next-to-leading order pQCD disagrees with experiment
both quantitatively and qualitatively. In contrast, we find that
the phenomenological model with off-shell non-collinear partons
successfully describes both the double differential Drell-Yan
cross section and the $p_T$ spectrum of Drell-Yan pairs without
the need of a $K$ factor.

The analysis of the Drell-Yan process cross section in our model
in the Drell-Yan scaling limit has shown that the phenomenological
model parametrizes
higher twist effects. Higher twist
contributions were up to date usually considered to be small,
because they are suppressed by powers of the hard scale. As a
rule, they are neglected in pQCD calculations. However, the power
suppressed effect can be large at realistic energies.

We have found that the intrinsic transverse momentum of quarks
generates both leading twist and $1/(\tau s)=1/M^2$ suppressed
effects. This is in line with our analysis of section~\ref{NLO},
which has shown that only part of observed $\langle p_T^2\rangle$
can be explained by NLO effects. In addition, we have shown that
next-to-leading twist corrections due to quark off-shellness lead
to an overall cross section suppression and are therefore
responsible for a part of the K-factor type discrepancy between
the leading order pQCD and data. Although we have shown this only
for a Breit-Wigner parametrization of the spectral function, we do
not expect these results to depend on its special form, since all
calculated cross sections involve only integrals over the spectral
functions.

If a Breit-Wigner parametrization for a quark spectra function is
used, the next-to-leading contribution is proportional to $\Gamma
/\sqrt{\tau s}$. This leads us to suggest that the quark spectral
function width $\Gamma$ scales as $\Gamma (M) \sim 1/M$ at large
hard scale $M=\sqrt{\tau s}$.

The formula that we obtained for the Drell-Yan cross section at
the next-to-leading twist level can be very useful for
applications, for example, in an event generator. Indeed, it
requires no numerical integration, while providing a good
approximation to the full calculations at $M \gtrsim 5$~GeV.
However, at $M \lesssim 5$~GeV, one has to go beyond the
next-to-leading order in the power series and use the formulas
of~\cite{paper2}.

The results show that the higher twist corrections to high energy
pro\-cesses can be large. Therefore, a detailed study and
modelling of these effects is necessary, if one hopes to reliably
extract quark and gluon properties from hadron scattering data.

%


\end{document}